# Comparing two methods of regularization of the kinetic energy density.

Dan Solomon
Rauland-Borg Corporation
Mount Prospect, IL
Email: dan.solomon@rauland.com
Aug. 9, 2011

**Abstract.**

In this paper we will compare two different methods of regularizing the kinetic energy density for a massless scalar field in the presence of a static scalar potential. One method of regularization is to subtract the cosmological constant from the *naïve* expression for the kinetic energy density. The other method is to use point split regularization. It is found that the two methods yield different results. The result obtained using point split regularization includes an extra ambiguous term.

## 1. Introduction.

In quantum field theory there is a potential problem involved with the calculation of kinetic energy density. This is because a *naïve* calculation, using the standard expression for the kinetic energy density operator, produces a divergent result. In order to produce a finite result the divergence must be removed. The process by which is done is called regularization.

It has been shown in [1] that there is a problem with "point split" regularization. It was shown that when this process is used an ambiguous term will show up in the result. In order to explore this problem in more detail we will calculate the kinetic energy density for a scalar field using two different methods of regularization and compare the results. The first method used will be to follow the approach in [2] and subtract off the cosmological constant to obtain an expression for the kinetic energy density. The second method will be that of point split regularization. It will be shown that two different methods produce different results.

The kinetic energy density will be determined for a zero mass scalar field in 1-1 dimensional space-time in the presence of an inverse square well scalar potential. In this case the field operator $\hat{\varphi}_\lambda(x,t)$ satisfies the Klein-Gordon equation,

$$\frac{\partial^2 \hat{\varphi}_\lambda}{\partial t^2} - \frac{\partial^2 \hat{\varphi}_\lambda}{\partial x^2} + \lambda V(x)\hat{\varphi}_\lambda = 0 \tag{1.1}$$

where $\lambda V(x)$ is the scalar potential and $\lambda$ is a non-negative constant that can be used to set the scalar potential to zero by specifying $\lambda = 0$. For the purposes of this discussion,

$$V(x) = \begin{cases} 1 \text{ for } |x| < a \\ 0 \text{ for } |x| > a \end{cases} \tag{1.2}$$

The kinetic energy density operator is defined by,

$$\hat{T}_{00}[\hat{\varphi}_\lambda] = \frac{1}{2}\left(\frac{\partial \hat{\varphi}_\lambda}{\partial t} \cdot \frac{\partial \hat{\varphi}_\lambda}{\partial t} + \frac{\partial \hat{\varphi}_\lambda}{\partial x} \cdot \frac{\partial \hat{\varphi}_\lambda}{\partial x}\right) \tag{1.3}$$

Let $|0_\lambda\rangle$ be a normalized state vector which will be specified in more detail in the next section. The kinetic energy density expectation value for this state vector is given by,

$$T_{00,\lambda}(x,t) = \langle 0_\lambda | \hat{T}_{00}[\hat{\varphi}_\lambda(x,t)] | 0_\lambda \rangle \tag{1.4}$$

It can easily be shown that this quantity is infinite. In order to get a finite result it is necessary to subtract off an infinite term. Therefore the "renormalized" kinetic energy density is given by,

$$T_{00R}(x,t) = T_{00,\lambda}(x,t) - T_r \tag{1.5}$$

where $T_r$ is an infinite renormalization constant.

Note that this is still a formal expression because we are subtracting one infinite term from another infinite term which is mathematically meaningless. In order to produce a well defined result we must specify the details on how this subtraction is done. In Section 2 we will follow the approach of Graham and Olum [2] and renormalize the kinetic energy density by subtracting off the cosmological constant. Then, in Section 3, we will re-calculate the kinetic energy density using the method of point split regularization. It will be shown that when this latter method is used an extra term appears in the final result. In Section 4 we will examine an alternative approach to calculating the kinetic energy density where regularization is not required. It is shown that this approach gives the same result as method used in Section 2 which is to subtract off the cosmological constant. This suggests that there is a flaw in the method of point split regularization.

## 2. Kinetic energy density 1.

In this section we will use results from [2] and [3] to obtain an expression for the renormalized kinetic energy density for a scalar field with zero mass in 1-1 dimensional space-time in the presence the inverse square well scalar potential given by (1.2). The field operator that satisfies (1.1) is given by,

$$\hat{\varphi}_\lambda(x,t) = \int_0^\infty \frac{dk}{\sqrt{2\pi\omega}} \sum_{\chi=+,-} \left( \hat{a}_{\lambda,k}^\chi f_{\lambda,k}^\chi(x,t) + \hat{a}_{\lambda,k}^{\chi*} f_{\lambda,k}^{\chi*}(x,t) \right) \qquad (2.1)$$

where,

$$f_{\lambda,k}^\chi(x,t) = \psi_{\lambda,k}^\chi(x) e^{-i\omega t} \qquad (2.2)$$

and where $\omega = |k|$ and the $\psi_{\lambda,k}^\chi(x)$ are real with $\psi_{\lambda,k}^-(x)$ representing the anti-symmetric solutions and $\psi_{\lambda,k}^+(x)$ representing the symmetric solutions. The $\hat{a}_{\lambda,k}^\chi$ and $\hat{a}_{\lambda,k}^{\chi*}$ are the annihilation and creation operators, respectively. They obey the commutation relationships $\left[ \hat{a}_{\lambda,k'}^{\chi'}, \hat{a}_{\lambda,k}^{\chi*} \right] = \delta_{\chi'\chi} \delta(k'-k)$ with all other commutations being zero. The $\psi_{\lambda,k}^\chi(x)$ are solutions of,

$$-k^2 \psi_{\lambda,k}^\chi(x) - \frac{d^2 \psi_{\lambda,k}^\chi(x)}{dx^2} + \lambda V(x) \psi_{\lambda,k}^\chi(x) = 0 \qquad (2.3)$$

where,

$$\int_{-\infty}^{+\infty} dx\, \psi_{\lambda,k}^+(x) \psi_{\lambda,k'}^-(x) = 0 \text{ and } \sum_{\chi=+,-} \int_{-\infty}^{+\infty} \psi_{\lambda,k}^\chi(x) \psi_{\lambda,k'}^\chi(x)\, dx = 2\pi\delta(k-k') \qquad (2.4)$$

The above relationships define the field operator. The functions $\psi_{\lambda,k}^\chi(x)$ are given in Appendix 1. The normalized state vector on which the field operators act will be designated by $|0_\lambda\rangle$ and obeys the relationship $\hat{a}_{\lambda,k}|0_\lambda\rangle = 0$.

Use the above relationships along with (1.3) in (1.4) to obtain,

$$T_{00,\lambda}(x) = \int_0^\infty \left( \frac{dk}{2\pi|k|} \right) \sum_{\chi=+,-} \frac{1}{2} \left[ \left| \frac{\partial f_{\lambda,k}^\chi(x,t)}{\partial t} \right|^2 + \left| \frac{\partial f_{\lambda,k}^\chi(x,t)}{\partial x} \right|^2 \right] dk \qquad (2.5)$$

Next, use (2.2) in the above to obtain,

$$T_{00,\lambda}(x) = \int_0^\infty \left(\frac{dk}{2\pi|k|}\right) \sum_{\chi=+,-} \frac{1}{2}\left[k^2 \psi_{\lambda,k}^\chi(x)^2 + \left(\frac{d\psi_{\lambda,k}^\chi(x)}{dx}\right)^2\right] dk \qquad (2.6)$$

This is the unrenormalized kinetic energy density. In order to regularize this solution we proceed as in [2] and introduce a counter term to compensate for the cosmological constant which is the kinetic energy density when the scalar potential is zero. This involves subtracting off the term,

$$T_{00,0}(x) = \int_0^\infty \left(\frac{dk}{2\pi|k|}\right) \sum_{\chi=+,-} \frac{1}{2}\left[k^2 \psi_{0,k}^\chi(x)^2 + \left(\frac{d\psi_{0,k}^\chi(x)}{dx}\right)^2\right] dk \qquad (2.7)$$

where we obtained this expression by setting $\lambda = 0$ in (2.6). Referring to Appendix 1 we have,

$$\psi_{0,k}^+(x) = \cos(kx) \text{ and } \psi_{0,k}^-(x) = \sin(kx) \qquad (2.8)$$

Therefore,

$$T_{00,0}(x) = \int_0^\infty \left(\frac{kdk}{2\pi}\right) \qquad (2.9)$$

At this point we have the expression,

$$T_{00R}(x) = \int_0^\infty \left(\frac{dk}{2\pi|k|}\right) \sum_{\chi=+,-} \frac{1}{2}\left[k^2 \psi_{0,k}^\chi(x)^2 + \left(\frac{d\psi_{0,k}^\chi(x)}{dx}\right)^2\right] - \int_0^\infty \left(\frac{kdk}{2\pi}\right) \qquad (2.10)$$

Note that this expression is still "formal" because each integral when evaluated separately produces an infinite result. So we are subtracting one infinite term from another which is mathematically undefined. What allows us to achieve a well defined result is to follow the approach of [2] and group both integrands under one integral. In this case the regularized kinetic energy density is,

$$T_{00R}(x) = \int_0^\infty \left(\frac{dk}{4\pi|k|}\right) \sum_{\chi=+,-} \left[\left(k^2 \psi_{\lambda,k}^\chi(x)^2 + \left(\frac{d\psi_{\lambda,k}^\chi(x)}{dx}\right)^2\right) - k^2\right] \qquad (2.11)$$

To show that this quantity will be finite use the results from Appendix 1. For $|x| > a$ we obtain

$$T_{00R}(|x| > a) = 0 \qquad (2.12)$$

and for $|x| < a$ we obtain,

$$T_{00R}(|x| < a) = \int_0^\infty T_k(x) dk \tag{2.13}$$

$$T_k(x) = \left(\frac{1}{4\pi |k|}\right) \left\{ \begin{array}{l} k^2 \left[ A_{+,k}^2 + A_{-,k}^2 - 2 \right] \\ -\lambda \left[ A_{+,k}^2 \sin^2\left(x\sqrt{k^2 - \lambda}\right) + A_{-,k}^2 \cos^2\left(x\sqrt{k^2 - \lambda}\right) \right] \end{array} \right\} \tag{2.14}$$

In the limit $k \to \infty$ $T_k(x)$ falls off faster than $1/k^2$ so that $T_{00R}(|x| < a)$ is finite. Therefore using this method of regularization we achieve a finite result for the renormalized kinetic energy density.

Using the above along with Appendix 1 to obtain,

$$T_{00R}(x)\Big|_{x<|a|} = \frac{\lambda}{2} \int_0^\infty \left(\frac{dk}{2\pi |k|}\right) \sum_{\chi=+,-} \left[ \left(\psi_{\lambda,k}^\chi(x)\right)^2 - \left(\psi_{\lambda,k}^\chi(-a)\right)^2 \right] \tag{2.15}$$

## 3. Kinetic energy density II.

In the last section we regularized the kinetic energy density by subtracting off the cosmological constant. In this section we will calculate the kinetic energy density using point split regularization. The point split regularization procedure is discussed in Chapter 2 of [4] and Section 4.6 of [5]. A portion of the following analysis is taken from Ref. [1].

Point split regularization is based on the observation that products of field operators are divergent if both operators are defined at a given point. However if the operators are defined at separate points the product with be finite. For example let $\hat{O}(x,t)$ stand for one of the following operators - $\hat{\varphi}(x,t)$, $\partial\hat{\varphi}(x,t)/\partial t$, or $\partial\hat{\varphi}(x,t)/\partial x$. Then $\langle 0_\lambda | \hat{O}(x,t)\hat{O}(x,t) | 0_\lambda \rangle$ is divergent but $\langle 0_\lambda | \hat{O}(x,t)\hat{O}(x',t') | 0_\lambda \rangle$ is finite as long as $(x,t) \neq (x',t')$. Therefore to give $\langle 0_\lambda | \hat{O}(x,t)\hat{O}(x,t) | 0_\lambda \rangle$ a mathematical meaning we can define it as $\langle 0_\lambda | \hat{O}(x,t)\hat{O}(x',t') | 0_\lambda \rangle$ in the limit that $(x',t') \to (x,t)$.

With this in mind we define the point split kinetic energy operator as follows. Define the following ordered pairs, $y = (y_0, y_1)$ and $y' = (y_0', y_1')$ where,

$$y_0 = t + \varepsilon_0/2; \; y_0' = t - \varepsilon_0/2; \; y_1 = x + \varepsilon_1/2; \; y_1' = x - \varepsilon_1/2 \tag{3.1}$$

The point split kinetic energy density operator is defined by,

$$\hat{T}_{00}(y,y';[\hat{\varphi}_\lambda]) = \frac{1}{2}\left(\hat{t}_{00}(y,y';[\hat{\varphi}_\lambda]) + \hat{t}_{00}(y',y;[\hat{\varphi}_\lambda])\right) \tag{3.2}$$

where,

$$\hat{t}_{00}(y,y';[\hat{\varphi}_\lambda]) = \frac{1}{2}\left(\frac{\partial\hat{\varphi}_\lambda(y)}{\partial y_0}\frac{\partial\hat{\varphi}_\lambda^*(y')}{\partial y'_0} + \frac{\partial\hat{\varphi}_\lambda(y)}{\partial y_1}\frac{\partial\hat{\varphi}_\lambda^*(y')}{\partial y'_1}\right) \tag{3.3}$$

Note that we have defined $\hat{T}_{00}(y,y';[\hat{\varphi}_\lambda])$ to be symmetric in $y$ and $y'$. The renormalized kinetic energy density expectation value is,

$$T_{00R}(x,t;\varepsilon) = \langle 0_\lambda|\hat{T}_{00}(y,y';[\hat{\varphi}_\lambda])|0_\lambda\rangle - \langle 0_0|\hat{T}_{00}(y,y';[\hat{\varphi}_0])|0_0\rangle \tag{3.4}$$

where the above expression is evaluated in the limit that the pair $\varepsilon = (\varepsilon_0,\varepsilon_1) \to 0$. Each of the quantities on the right of the equal sign are finite as long as $(\varepsilon_0,\varepsilon_1) \neq 0$.

Obviously the quantity $T_{00R}(x,t;\varepsilon)$ is dependent on the pair $\varepsilon = (\varepsilon_0,\varepsilon_1)$. If point split regularization works then the terms containing $(\varepsilon_0,\varepsilon_1)$ should not appear in $T_{00R}(x,t;\varepsilon)$ in the limit that $(\varepsilon_0,\varepsilon_1) \to 0$.

Using (2.1) in (3.2) we obtain,

$$\langle 0_\lambda|\hat{T}_{00}(y,y';[\hat{\varphi}_\lambda])|0_\lambda\rangle = \int_0^\infty \sum_{\chi=\pm} \xi_{\lambda,k}^\chi(y,y')\,dk \tag{3.5}$$

where,

$$\xi_{\lambda,k}^\chi(y;y') = \frac{1}{4\pi|k|}\left[\left(\frac{\partial f_{\lambda,k}^\chi(y)}{\partial y_0}\frac{\partial f_{\lambda,k}^{\chi*}(y')}{\partial y'_0} + \frac{\partial f_{\lambda,k}^\chi(y)}{\partial y_1}\frac{\partial f_{\lambda,k}^{\chi*}(y')}{\partial y'_1}\right) + c.c.\right] \tag{3.6}$$

$\xi_{\lambda,k}^\chi(y;y')$ is the point split kinetic energy density of the $\chi k - th$ mode. In order to make sure that the integrals are well defined we will add a frequency cutoff factor $e^{-k\tau}$ to obtain,

$$T_{00,\lambda}(x,t;\varepsilon,\tau) = \int_0^\infty \xi_{\lambda,k}(y,y')e^{-k\tau}\,dk \tag{3.7}$$

where $\tau \to 0$ and,

$$\xi_{\lambda,k}(y;y') = \sum_{\chi=\pm} \xi_{\lambda,k}^\chi(y;y') \tag{3.8}$$

and where we have substituted $T_{00,\lambda}(x,t;\varepsilon,\tau)$ for $\langle 0_\lambda|\hat{T}_{00}(y,y';[\hat{\varphi}_\lambda])|0_\lambda\rangle$.

Set $\lambda = 0$ in (3.7) to obtain,

$$T_{00,0}(x,t;\varepsilon,\tau) = \int_0^\infty \xi_{0,k}(y, y') e^{-k\tau} dk \qquad (3.9)$$

Therefore the renormalized kinetic energy density is,

$$T_{00R}(x,t;\varepsilon,\tau) = T_{00,\lambda}(x,t;\varepsilon,\tau) - T_{00,0}(x,t;\varepsilon,\tau) \qquad (3.10)$$

Use (3.7) and (3.9) in the above to obtain,

$$T_{00R}(y; y') = \int_0^\infty \left( \xi_{\lambda,k}(y; y') - \xi_{0,k}(y; y') \right) e^{-k\tau} dk \qquad (3.11)$$

The kinetic energy density in the region $|x| < a$ will be calculated and the result will be compared to (2.11). From Appendix 2, for this region, we evaluate $\xi_{\lambda,k}(y; y')$ and $\xi_{0,k}(y; y')$ to obtain,

$$\xi_{\lambda,k}(x,t;\varepsilon_0,\varepsilon_1) = \left[ \left( \frac{\cos(k\varepsilon_0)}{4\pi} \right) \begin{pmatrix} \left( A_{+,k}^2 + A_{-,k}^2 \right) \left( k - \frac{\lambda}{2k} \right) \cos\left( \varepsilon_1 \sqrt{k^2 - \lambda} \right) \\ + \left( A_{-,k}^2 - A_{+,k}^2 \right) \frac{\lambda}{2k} \cos\left( 2x\sqrt{k^2 - \lambda} \right) \end{pmatrix} \right] \qquad (3.12)$$

Set $\lambda = 0$ in the above to obtain,

$$\xi_{0,k}(x,t;\varepsilon_0,\varepsilon_1) = \left( \frac{\cos(k\varepsilon_0)}{2\pi} \right) k \cos(k\varepsilon_1) \qquad (3.13)$$

Define the quantities,

$$R_k(x,t;\varepsilon_0,\varepsilon_1) = \frac{\lambda}{2\pi} \varepsilon_1 \cos(k\varepsilon_0) \sin(k\varepsilon_1) \qquad (3.14)$$

and,

$$S_k(x,t;\varepsilon_0,\varepsilon_1) = \left( \xi_{\lambda,k}(x,t;\varepsilon_0,\varepsilon_1) - \xi_{0,k}(x,t;\varepsilon_0,\varepsilon_1) \right) - R_k(x,t;\varepsilon_0,\varepsilon_1) \qquad (3.15)$$

Use the above in (3.11) to obtain,

$$T_{00R}(x,t;\varepsilon_0,\varepsilon_1,\tau) = \int_0^\infty S_k(x,t;\varepsilon_0,\varepsilon_1) e^{-k\tau} dk + \int_0^\infty R_k(x,t;\varepsilon_0,\varepsilon_1) e^{-k\tau} dk \qquad (3.16)$$

To evaluate the above quantity we start by writing,

$$\int_0^\infty S_k(x,t;\varepsilon_0,\varepsilon_1) e^{-k\tau} dk = \int_0^b S_k(x,t;\varepsilon_0,\varepsilon_1) e^{-k\tau} dk + \int_b^\infty S_k(x,t;\varepsilon_0,\varepsilon_1) e^{-k\tau} dk \qquad (3.17)$$

where the limit of integration, $b$, is given by,

$$b = \begin{cases} 1/\varepsilon_0^{(1/4)} \text{ if } \varepsilon_0 > \varepsilon_1 \\ 1/\varepsilon_1^{(1/4)} \text{ if } \varepsilon_1 > \varepsilon_0 \end{cases} \tag{3.18}$$

Next evaluate the quantity,

$$C_\lambda(x,t;\varepsilon_0,\varepsilon_1,\tau) = \int_0^b \xi_{\lambda,k}(x,t;\varepsilon_0,\varepsilon_1) e^{-k\tau} dk \tag{3.19}$$

With this restriction on the upper limit of integration we have,

$$\cos(k\varepsilon_0) \underset{\varepsilon_0 \to 0}{\cong} 1 - \frac{(k\varepsilon_0)^2}{2} \tag{3.20}$$

and,

$$\cos\left(\varepsilon_1 \sqrt{k^2 - \lambda}\right) \underset{\varepsilon_1 \to 0}{\cong} 1 - \frac{\varepsilon_1^2 (k^2 - \lambda)}{2} \tag{3.21}$$

Use this (3.19) to obtain,

$$C_\lambda(x,t;\varepsilon_0,\varepsilon_1,\tau) \underset{\varepsilon_0,\varepsilon_1 \to 0}{=} \int_0^b \xi_{\lambda,k}(x,t;0,0) e^{-k\tau} d\omega \tag{3.22}$$

where, from (3.12),

$$\xi_{\lambda,k}(x,t;0,0) = \left(\frac{1}{4\pi}\right) \left( \begin{array}{c} \left(A_{+,k}^2 + A_{-,k}^2\right)\left(k - \dfrac{\lambda}{2k}\right) \\ +\left(A_{-,k}^2 - A_{+,k}^2\right)\dfrac{\lambda}{2k}\cos\left(2x\sqrt{k^2 - \lambda}\right) \end{array} \right) \tag{3.23}$$

The reason for this result is that when (3.20) and (3.21) are used then the integral in (3.19) contains terms that go as $k^3 \varepsilon_1^2$, for example. This integrates out to $k^4 \varepsilon_1^2$. When the upper limit of integration, $b$, is substituted into this term we obtain $\left[\varepsilon_1^2 / (\varepsilon_1 \text{ or } \varepsilon_0)\right] \to 0$ as $(\varepsilon_0, \varepsilon_1) \to 0$.

It is easy to show that,

$$\int_0^b R_k(x,t;\varepsilon_0,\varepsilon_1) e^{-k\tau} dk \underset{\varepsilon_0,\varepsilon_1 \to 0}{=} 0 \tag{3.24}$$

It is shown in Appendix 3 that,

$$\int_b^\infty S_{\lambda,k}(x,t;\varepsilon_0,\varepsilon_1) e^{-k\tau} dk \underset{\varepsilon_0,\varepsilon_1 \to 0}{=} 0 \tag{3.25}$$

Therefore,

$$\int_0^\infty S_k(x,t;\varepsilon_0,\varepsilon_1)e^{-k\tau}dk \underset{\varepsilon_0,\varepsilon_1\to 0}{=} \int_0^b \left(\xi_{\lambda,k}(x,t;0,0)-\xi_{0,k}(x,t;0,0)\right)e^{-k\tau}dk \qquad (3.26)$$

where $\xi_{0,k}(x,t;0,0)$ is obtained from (3.23) by setting $\lambda = 0$.

$$\xi_{0,k}(x,t;0,0) = \frac{k}{2\pi} \qquad (3.27)$$

Referring to (2.14) we can show that,

$$T_k(x) = \xi_{\lambda,k}(x,t;0,0) - \xi_{0,k}(x,t;0,0) \qquad (3.28)$$

As was discussed in Section 2 it can be shown that for $k \to \infty$, $T_k(x)$ falls off faster than $1/k^2$. This means that,

$$\int_b^\infty \left(\xi_{\lambda,k}(x,t;0,0)-\xi_{0,k}(x,t;0,0)\right)e^{-k\tau}dk \underset{\varepsilon_0,\varepsilon_1\to 0}{=} 0 \qquad (3.29)$$

The result of all this that in the limit $\varepsilon_0, \varepsilon_1, \tau \to 0$ we obtain,

$$\int_0^\infty S_k(x,t;\varepsilon_0,\varepsilon_1)e^{-k\tau}dk \underset{\varepsilon_0,\varepsilon_1,\tau\to 0}{=} \int_0^\infty \left(\xi_{\lambda,k}(x,t;0,0)-\xi_{0,k}(x,t;0,0)\right)dk = T_{00R}(|x|<a) \qquad (3.30)$$

We have dropped the $e^{-k\tau}$ term on the right of the equals sign due to the fact that the integrand falls off sufficiently fast as $k \to \infty$.

Therefore in the region $|x|<a$ and in the limit $\varepsilon_0, \varepsilon_1, \tau \to 0$ we obtain,

$$T_{00R}(x,t;\varepsilon_0,\varepsilon_1,\tau) = T_{00R}(|x|<a) + \int_0^\infty R_k(x,t;\varepsilon_0,\varepsilon_1)e^{-k\tau}dk \qquad (3.31)$$

Recall that $T_{00R}(|x|<a)$ is the kinetic energy density as calculated in Section 2 by subtracting off the cosmological constant. Therefore the kinetic energy density $T_{00R}(x,t;\varepsilon_0,\varepsilon_1,\tau)$ as calculated by point split regularization includes an extra term which is the integral on the right side of the above expression. This term is readily evaluated as,

$$\int_0^\infty R_k(y;y')e^{-k\tau}dk = \frac{\lambda}{4\pi}\left[\left(\frac{\varepsilon_1^2}{\varepsilon_1^2-\varepsilon_0^2-2i\tau\varepsilon_0+\tau^2}\right)+\text{c.c.}\right] \qquad (3.32)$$

When we take the limit $(\varepsilon_0, \varepsilon_1, \tau) \to 0$ we find that this expression depends on the way in which the quantities $(\varepsilon_0, \varepsilon_1, \tau)$ approach zero. For example let's set $\tau = 0$. Then if $\varepsilon_0$ goes to zero much faster than $\varepsilon_1$, i.e. $\varepsilon_1 \gg \varepsilon_0 \to 0$, we obtain,

$$\int_0^\infty R_k(y; y') e^{-k\tau} dk = \frac{\lambda}{2\pi} \tag{3.33}$$

On the other hand if $\varepsilon_0 = \varepsilon_1 \to 0$ and $\tau = 0$ then,

$$\int_0^\infty R_k(y; y') e^{-k\tau} dk = \infty \tag{3.34}$$

Therefore this extra term is ambiguous. Unless we know how to take $(\varepsilon_0, \varepsilon_1, \tau)$ to zero, we cannot evaluate the kinetic energy density using point split regularization.

## 4. Another approach.

It has been shown that two different methods of regularization yield different results. Which one is correct? In order to resolve this issue we will consider a method to calculate the kinetic energy density where regularization is not required. To understand how this might be achieved recall that the problem with calculating the kinetic energy density is that the "formal" result is infinite. An infinite term must be subtracted out to get a physically meaningful result.

The reason this problem occurs was discussed at the beginning of Section 3. Quantities of the form $\langle 0_\lambda | \hat{O}(x,t) \hat{O}(x,t) | 0_\lambda \rangle$ are infinite and need to be regularized in order to be physically meaningful. On the other hand $\langle 0_\lambda | \hat{O}(x,t) \hat{O}(x',t') | 0_\lambda \rangle$ is finite for $(x,t) \neq (x',t')$. This suggests that even though $\langle 0_\lambda | \hat{O}(x,t) \hat{O}(x,t) | 0_\lambda \rangle$ may not be mathematically well-defined the first derivate of this with respect to $x$ or $t$ may be well-defined. Consider,

$$\frac{d}{dx} \langle 0_\lambda | \hat{O}(x,t) \hat{O}(x,t) | 0_\lambda \rangle = \langle 0_\lambda | \left( \frac{d\hat{O}(x,t)}{dx} \hat{O}(x,t) + \hat{O}(x,t) \frac{d\hat{O}(x,t)}{dx} \right) | 0_\lambda \rangle \tag{4.1}$$

Using the usual definition of the derivative we have,

$$\frac{d\hat{O}(x,t)}{dx} \underset{\varepsilon \to 0}{=} \frac{\hat{O}(x+\varepsilon, t) - \hat{O}(x-\varepsilon, t)}{2\varepsilon} \tag{4.2}$$

Using this in (4.1) we end up with terms of the form $\langle 0_\lambda | \hat{O}(x \pm \varepsilon, t) \hat{O}(x,t) | 0_\lambda \rangle$ and $\langle 0_\lambda | \hat{O}(x,t) \hat{O}(x \pm \varepsilon, t) | 0_\lambda \rangle$. All these terms will be mathematically well-defined for $\varepsilon \neq 0$.

With this in mind consider the formal expression for the kinetic energy density $T_{00,\lambda}(x)$ as given by Eq. (2.6). This quantity is infinite and must be regularized to produce a finite physically meaningful result. Next consider the first derivative of this quantity,

$$\frac{d}{dx} T_{00,\lambda}(x) = \int_0^\infty \left( \frac{dk}{2\pi |k|} \right) \sum_{\chi=+,-} \left[ \left( k^2 \psi_{\lambda,k}^\chi + \frac{d^2 \psi_{\lambda,k}^\chi}{dx^2} \right) \frac{d\psi_{\lambda,k}^\chi}{dx} \right] \quad (4.3)$$

Use (2.3) in the above to obtain,

$$\frac{d}{dx} T_{00,\lambda}(x) = \lambda V(x) \int_0^\infty \left( \frac{dk}{2\pi |k|} \right) \sum_{\chi=+,-} \left[ \psi_{\lambda,k}^\chi(x) \frac{d\psi_{\lambda,k}^\chi}{dx} \right] \quad (4.4)$$

Use (1.2) and Appendix 1 to obtain,

$$\frac{d}{dx} T_{00,\lambda}(x) \Big|_{|x|>a} = 0 \quad (4.5)$$

and,

$$\frac{d}{dx} T_{00,\lambda}(x) \Big|_{|x|<a} = \lambda \int_0^\infty \left( \frac{\sqrt{k^2 - \lambda} \, dk}{2\pi k} \right) \left( A_{-,k}^2 - A_{+,k}^2 \right) \cos\left(x\sqrt{k^2 - \lambda}\right) \sin\left(x\sqrt{k^2 - \lambda}\right) \quad (4.6)$$

In the limit that $k \to \infty$, $\left( A_{-,k}^2 - A_{+,k}^2 \right) \to 1/k^2$. Therefore the above integral is finite. The result is that even though $T_{00,\lambda}(x)$ is infinite and not well-defined mathematically the first derivate $dT_{00,\lambda}(x)/dx$ is finite and well-defined.

Refer to (1.5) to obtain, $T_{00,\lambda}(x) = T_{00R}(x) + T_r$ where $T_r$ is a constant. Use this in (4.4) to obtain,

$$\frac{d}{dx} T_{00R}(x) = \lambda V(x) \int_0^\infty \left( \frac{dk}{2\pi |k|} \right) \sum_{\chi=+,-} \left[ \psi_{\lambda,k}^\chi(x) \frac{d\psi_{\lambda,k}^\chi}{dx} \right] \quad (4.7)$$

Therefore we have obtained a finite and physically meaningful expression for the first derivative of the renormalized kinetic energy density. We have done this without having to go through a "regularization" procedure. That is, we don't have to subtract off an infinite term and we can work with the expression as is.

Next, consider $T_{00R}(x)$ in the region $|x| > a$. Referring to (4.7) and (1.2) $dT_{00R}(x)/dx = 0$ in this region so that $T_{00R}(x)$ is constant for $|x| > a$. What is the value of this constant. The total kinetic energy is the integration of $T_{00R}(x)$ over all space. Therefore if $T_{00R}(x) \ne 0$ in the region $|x| > a$ the total kinetic energy will be infinite. We will make the assumption that the total kinetic energy is finite. Therefore $T_{00R}(x) = 0$ in the region $|x| > a$. This allows us to evaluate $T_{00R}(x)$ in the region $|x| < a$ by integrating (4.7) to obtain,

$$T_{00R}(x)\underset{x \le |a|}{=} \frac{\lambda}{2} \int_0^\infty \left( \frac{dk}{2\pi |k|} \right) \sum_{\chi=+,-} \left[ \left(\psi_{\lambda,k}^\chi(x)\right)^2 - \left(\psi_{\lambda,k}^\chi(-a)\right)^2 \right] \qquad (4.8)$$

This the same as the result we obtained in Section 2 (see Eq. (2.15)) where regularization was achieved by subtracting out the cosmological constant.

## 5. Conclusion.

We have used two different methods of renormalization to calculate the kinetic energy density. The first method was to subtract the cosmological constant from the *naïve* expression for the kinetic energy density. This resulted in Eq. (2.11) which was shown to be finite. The second method was to use point split regularization. In this case the kinetic energy density was evaluated in the region $|x| < a$. This resulted in Eq. (3.31) which included an extra term which is ambiguous. Finally, in Section 4 we used a method that did not require an explicit regularization step. The results of this method agreed with those of Section 2 where the regularization was achieved by subtracting out the cosmological constant. The conclusion is that the method of point split regulation is not a satisfactory method of regularization for this type of problem.

## Appendix 1.

We will determine the mode solutions $\psi_{\lambda,k}^{\chi}(x)$. These have been previously derived in Ref. [1]. However in order to set the notation and make the paper self-contained they are re-derived below.

The $\psi_{\lambda,k}^{\chi}(x)$ must satisfy (2.3). Start with the symmetric solutions with $\chi = +$. For the region where $|x| < a$ we have,

$$\psi_{\lambda,k}^{+}(x)\Big|_{|x|<a} = A_{+,k} \cos\left(x\sqrt{k^2 - \lambda}\right) \tag{5.1}$$

For $|x| > a$ we have,

$$\psi_{\lambda,k}^{+}(x)\Big|_{|x|>a} = \cos\left(k|x| + \delta_{+,k}\right) \tag{5.2}$$

For the anti-symmetric solutions ($\chi = -$) for $|x| < a$ we have,

$$\psi_{\lambda,k}^{-}(x)\Big|_{|x|<a} = A_{-,k} \sin\left(x\sqrt{k^2 - \lambda}\right) \tag{5.3}$$

For $|x| > a$ we have,

$$\psi_{\lambda,k}^{-}(x)\Big|_{|x|>a} = \sin\left(kx + \delta_{-,k}\varepsilon(x)\right) \tag{5.4}$$

where $\varepsilon(x) = +1$ for $x > 0$ and $\varepsilon(x) = -1$ for $x < 0$.

We solve for the $A_{\chi,k}$ use the boundary conditions at $x = \pm a$. These are that $\psi_{\lambda,k}^{\chi}(x)$ along with the first derivative $d\psi_{\lambda,k}^{\chi}(x)/dx$ are continuous across the boundary. Use this to obtain for the symmetric solutions,

$$\cos(ka + \delta_{+,k}) = A_{+,k} \cos\left(a\sqrt{k^2 - \lambda}\right) \tag{5.5}$$

and,

$$-k\sin(ka + \delta_{+,k}) = -A_{+,k}\sqrt{k^2 - \lambda}\sin\left(a\sqrt{k^2 - \lambda}\right) \tag{5.6}$$

These can be solved to obtain,

$$A_{+,k}^2 = \frac{1}{\left[1 - \frac{\lambda}{k^2}\sin^2\left(a\sqrt{(k^2 - \lambda)}\right)\right]} \tag{5.7}$$

Similarly for the anti-symmetric solutions we obtain,

$$\sin(ka+\delta_{-,k}) = A_{-,k} \sin\left(a\sqrt{k^2-\lambda}\right) \tag{5.8}$$

and

$$k\cos(ka+\delta_{-,k}) = A_{-,k} \sqrt{k^2-\lambda} \cos\left(a\sqrt{k^2-\lambda}\right) \tag{5.9}$$

These yield,

$$A_{-,k}^2 = \frac{1}{\left[1 - \frac{\lambda}{k^2} \cos^2\left(a\sqrt{(k^2-\lambda)}\right)\right]} \tag{5.10}$$

In the above expressions $k$ takes on the values 0 to $\infty$. In the case that $k < \lambda$ we have $\sqrt{(k^2-\lambda)} = i\sqrt{(\lambda-k^2)}$.

## Appendix 2.

We will calculate the kinetic energy density $\xi_{\lambda,k}^\chi(y;y')$ of a given mode in the region $|x| < a$. First use (2.2) in (3.6) to obtain,

$$\xi_{\lambda,k}^\chi(y;y') = \left[\left(\frac{e^{ik(y_0-y_0')}}{4\pi k}\right)\left(k^2 \psi_{\lambda,k}^\chi(y_1)\psi_{\lambda,k}^{\chi*}(y_1') + \frac{\partial \psi_{\lambda,k}^\chi(y_1)}{\partial y_1}\frac{\partial \psi_{\lambda,k}^{\chi*}(y_1')}{\partial y_1'}\right)\right] + c.c \tag{6.1}$$

Next refer to Appendix 1 to obtain,

$$\xi_{\lambda,k}^+(y;y') = \left[A_{+,k}^2 \left(\frac{e^{ik(y_0-y_0')}+c.c}{4\pi k}\right)\begin{pmatrix}k^2\cos\left((y_1-y_1')\sqrt{k^2-\lambda}\right) \\ -\lambda\sin\left(y_1\sqrt{k^2-\lambda}\right)\sin\left(y_1'\sqrt{k^2-\lambda}\right)\end{pmatrix}\right] \tag{6.2}$$

and,

$$\xi_{\lambda,k}^-(y;y') = \left[A_{-,k}^2 \left(\frac{e^{ik(y_0-y_0')}+c.c}{4\pi k}\right)\begin{pmatrix}k^2\cos\left((y_1-y_1')\sqrt{k^2-\lambda}\right) \\ -\lambda\cos\left(y_1\sqrt{k^2-\lambda}\right)\cos\left(y_1'\sqrt{k^2-\lambda}\right)\end{pmatrix}\right] \tag{6.3}$$

Use this in Eq. (3.8) to obtain,

$$\xi_{\lambda,k}(y;y') = \left[\left(\frac{\cos(k(y_0-y_0'))}{4\pi}\right)\begin{pmatrix}(A_{+,k}^2+A_{-,k}^2)\left(k-\frac{\lambda}{2k}\right)\cos\left((y_1-y_1')\sqrt{k^2-\lambda}\right) \\ +(A_{-,k}^2-A_{+,k}^2)\frac{\lambda}{2k}\cos\left((y_1+y_1')\sqrt{k^2-\lambda}\right)\end{pmatrix}\right]$$

Use (3.1) in the above to obtain (3.12) in the text.

## Appendix 3.

In the limit that $k \gg \lambda$ we can write,

$$A_{+,k}^2 = 1 + \frac{\lambda}{k^2}\sin^2\left(a\sqrt{(k^2-\lambda)}\right) + O(1/k^2) \tag{7.1}$$

and,

$$A_{-,k}^2 = 1 + \frac{\lambda}{k^2}\cos^2\left(a\sqrt{(k^2-\lambda)}\right) + O(1/k^4) \tag{7.2}$$

$$\cos\left(\varepsilon_1\sqrt{k^2-\lambda}\right) \underset{k \gg \lambda}{=} \cos(k\varepsilon_1) + \frac{\lambda\varepsilon_1}{k}\sin(k\varepsilon_1) - \frac{\lambda^2\varepsilon_1^2}{8k^2}\cos(k\varepsilon_1) + O(1/k^3) \tag{7.3}$$

$$\xi_{\lambda,k}(x,t;\varepsilon_0,\varepsilon_1) \underset{k \gg \lambda}{=} \left(\frac{\cos(k\varepsilon_0)}{4\pi}\right)\left(2k\left(1-\frac{\lambda^2}{4k^4}\right)\left(\begin{array}{c}\cos(k\varepsilon_1) + \frac{\lambda\varepsilon_1}{k}\sin(k\varepsilon_1) \\ -\frac{\lambda^2\varepsilon_1^2}{8k^2}\cos(k\varepsilon_1) + O(1/k^3)\end{array}\right)\right) + O(1/k^3) \tag{7.4}$$

Therefore, in the limit $k \gg \lambda$ and $\varepsilon_0, \varepsilon_1 \to 0$,

$$\int_b^\infty \left(\xi_{\lambda,k}(x,t;\varepsilon_0,\varepsilon_1) - \xi_{0,k}(x,t;\varepsilon_0,\varepsilon_1)\right) e^{-k\tau} dk = \int_b^\infty \frac{\lambda\varepsilon_1}{2\pi}\sin(k\varepsilon_1)\cos(k\varepsilon_0) \tag{7.5}$$

Refer to (3.14) to obtain (3.25) in the text.

## References.